# Towards Biochemical Filter with Sigmoidal Response to pH Changes: Buffered Biocatalytic Signal Transduction

Marcos Pita[#], Vladimir Privman[‡], Mary A. Arugula[‡],
Dmitriy Melnikov[‡], Vera Bocharova[‡], Evgeny Katz[‡]*

[#] *Instituto de Catálisis y Petroleoquímica, CSIC, C/Marie Curie 2, 28040 Madrid, Spain*
[‡] *Department of Chemistry and Biomolecular Science, and Department of Physics, Clarkson University, Potsdam, NY 13699, USA*

---

**\*** Corresponding author:
E-mail: ekatz@clarkson.edu; Tel.: +1 (315) 268-4421; Fax: +1 (315) 268-6610

**Abstract**

We realize a biochemical filtering process by introducing a buffer in a biocatalytic signal-transduction logic system based on the function of an enzyme, esterase. The input, ethyl butyrate, is converted into butyric acid—the output signal, which in turn is measured by the drop in the pH value. The developed approach offers a versatile "network element" for increasing the complexity of biochemical information processing systems. Evaluation of an optimal regime for quality filtering is accomplished in the framework of a kinetic rate-equation model.

1.      Introduction

As silicon technology is approaching its limits,[1] unconventional approaches to the next generation computing systems are being researched with the hope of offering new functionalities and advances in information processing.[2,3] Molecular (chemical) computing[4-10] has been considered among the approaches to miniaturizing computing elements, as well as novel applications. Biomolecular computing[11-13] can offer an additional advantage of the biochemical specificity of catalytic and recognition processes, ultimately aiming at mimicking and developing systems compatible with the natural information processing mechanisms. Biochemical systems designed for information processing range from various biomolecules, such as proteins,[14,15] DNA,[16] RNA,[17] DNAzymes,[18,19] to whole biological cells operating as computing devices.[20] Enzyme-based biocatalytic systems realizing binary logic gates[11,21-24] and their small networks[25,26] have been recently extensively studied in biomolecular computing.

Despite great expectations for biomolecular computing (biocomputing) systems,[27] the present level of their complexity does not allow any real computing device based on biomolecules. Indeed, only networks performing a few logic operations on a time-scale of minutes have thus far been realized in the lab. However, another application for (bio)molecular information processing has been within reach for the available technology level: extension of capabilities of multi-signal digital biosensors with built-in logic.[28] Such biosensors processing information at the biochemical level are of interest in biomedical applications,[29-32] since biomolecules are capable of operating in a biological environment.[33] Within the general program of the digital biosensor development, several systems of various complexity have recently been designed to analyze pathophysiological conditions corresponding to different injuries.[34-38]

Another promising application of biomolecular logic systems has been for controlling multi-signal responsive materials aiming at chemical actuators with built-in logic.[39] Coupling of the signal-processing enzyme logic systems with switchable "smart" materials can be achieved through redox transformations[40] or pH changes[41-45] driven by the enzyme reactions, offering new "Sense/Act" sensor/actuator functionalities. Specifically, pH changes generated by the enzyme logic systems, causing polymer materials to switch between different states, have been successfully used to trigger reconfiguration of various nanostructured systems such as membranes,[41] emulsions[42] and nanoparticle assemblies,[43] as well as modified electrodes[46] and bioelectronic devices (biofuel cells).[47]

One of the main challenges for biocomputing has been the design of logic gates that can be combined, with the help of other, enabling non-binary elements, to allow interconnection in fault-tolerant networks with control of noise buildup, for information processing of increasing size and complexity.[48-50] There is ample experimental evidence[48-50] that the level of noise in (bio)chemical computing systems is quite high compared to their electronic counterparts: This includes noise in both the inputs/outputs chemical and the enzyme concentrations which typically vary at least several percent. Thus, the problem of noise amplification and its control becomes an important issue in the design of even small biocomputing networks.[51,52]

One of the possible approaches to noise reduction and control could be the use of filters as network elements converting a convex-shape concentration-response typical of (bio)chemical reactions to a sigmoidal function, thus suppressing noise at the binary **0** and **1** logic points. This could be achieved, for example, by using enzymes with substrates that have self-promoter properties, as biocatalytic elements in logic gates/networks.[53,54] However, this approach requires very specific (e.g., certain allosteric) enzymes and thus cannot be considered as a versatile solution. We have recently reported a general approach to biochemical filter systems based on redox transformations.[55] Since many enzyme logic systems use pH changes as output signals to control pH-responsive materials[41-44] and switchable electrode interfaces,[46,47] an alternative approach devised specifically for pH-change signal filtering would be desirable. The present paper reports the first experimental realization and theoretical modeling of a versatile pH-filter mechanism based on buffering, for enzyme-catalyzed reactions, aiming at noise reduction upon transduction of biochemical signals.

Our system is illustrated in Figure 1. The "logic function" considered is the simplest possible one, that of signal transmission/transduction/conversion: an enzyme-catalyzed reaction converts the concentration of the input chemical into the output signal quantified by the drop in the pH value. An effect of a proper quantity of buffer, if added, is to change the system response from convex, as typical for most (bio)chemical processes, to sigmoidal, as desired for filters. Details of our experimental system are presented in Sections 2-3. Careful selection of the system parameters by modeling is crucial for obtaining a reasonable filtering effect for suppression of noise buildup, as explained in Sections 4-5. Section 6 is devoted to conclusions.

## 2. Experimental

Esterase from porcine liver (EC 3.1.1.1), ethyl butyrate, 4-(2-hydroxyethyl)-1-piperazineethanesulfonic acid (HEPES buffer), and other inorganic reagents were purchased from Sigma-Aldrich and were used as supplied without any pretreatment or purification. Ultrapure water (18.2 MΩ·cm) from NANOpure Diamond (Barnstead) source was used in all of the experiments.

The enzymatic reactions were carried out in aqueous solutions containing 0 mM, 50 mM and 100 mM of HEPES buffer and esterase (4 U·mL$^{-1}$). Prior to starting each reaction, the pH of the buffer was adjusted to 7.0 by using 0.1 M NaOH. Once the pH was stable, the reaction was started by adding ethyl butyrate as input with the concentration ranging from 0.1 mM to 100 mM and monitoring the decrease in the pH value. The experiments were performed under vigorous stirring in a final solution volume of 5 mL. The pH measurements were performed with Mettler Toledo® SevenEasy pH-meter. Using Lab-pH software, the readout of the pH was performed every second. The decrease of pH was measured for the time period of 120 min. Additionally control experiments with butyric acid varied from 0.1 mM to 100 mM, with 0 mM, 50 mM, 100 mM of HEPES buffer were also performed. All experiments were performed at an ambient temperature of 23±2 °C.

## 3. Biochemical pH-Signal Filter

Our enzymatic biochemical "signal processing" is based on the hydrolysis of ethyl butyrate (substrate, S, serving as the logic input) catalyzed by esterase (enzyme, E) with the butyric acid as the logic output product (P), causing the drop in the pH value. Schematically, we write

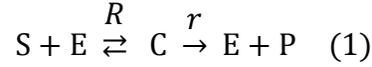

$$S + E \underset{}{\overset{R}{\rightleftarrows}} C \overset{r}{\rightarrow} E + P \quad (1)$$

where $R$ and $r$ are the forward rates at which the intermediate complex, C, and the final product are formed (the first step can be reversible: we comment on this later). While the output product is the acid molecule,[56,57] we assume that the equilibrium dissociation of butyric acid (A) in solution, with the dissociation constant $K_A$, is instantaneous: we use the self-explanatory notation

$$[P] = [A] + [A^-]$$
$$K_A = [A^-][H^+]/[A] \quad (2)$$

It is well known[58-60] that the reaction rates $R$ and $r$ for esterase strongly depend on pH of the system: The reaction (1) is fastest at pH $\sim$ 9 and it virtually stops at pH $\sim$ 3. To account for the slowing down of the reaction with decreasing pH, in our parameter range we can assume[58] acidic ionization of the active sites in both the enzyme and intermediate complex, by adding instantaneous dissociation equilibria

$$K_E = [E][H^+]/[EH^+]$$
$$K_C = [C][H^+]/[CH^+] \quad (3)$$

for the ionized enzyme and complex, respectively. We monitor the pH value of the solution as a function of time, $t$, starting from the reaction's on-set ($t = 0$). This yields kinetic data for the reaction which will be used in the theoretical model to determine the reaction constants and perform system optimization.

We could consider zero initial concentrations of the input substrate and the output product as the logic-**0** values, while the case of the maximum initial substrate concentration and the output product measured at a specific reaction time $t$ — which in practice are set by an application — as corresponding to logic-**1** points. Another option, favored in our present study and used from now on, is to set the logic values for the output at the appropriate pH values. The "logic range" variables x and y (inset in Figure 1) that vary between 0 and 1, will be defined in the next section. It is important to note that beyond such a "binary" description, in general the plot of the output vs. the initial substrate concentration (viewed as the response curve in terms of the logic-range variable of the type shown in Figure 1) is convex, which is typical for enzymatic reactions; such response amplifies input analog noise.[48-50] Instead, it is desirable to have a sigmoidal response, i.e., the curve should be "flat" around both logic points and with inflection in between. Indeed, such a response curve offers the filtering effect by decreasing analog noise at each logic point.[55]

While such a sigmoidal response is observed in nature,[50,61-63] it is generally not easy to realize in biochemical reactions.[52,55] In this work, we attempt to artificially create sigmoidal response of the pH variation in an enzyme-catalyzed reaction: We introduce, see Figure 1, a buffer (B), here HEPES. The latter is a weak acid with the dissociation constant satisfying $K_{\text{HEPES}} \ll K_A$,

$$K_{\text{HEPES}} = [\text{B}^-][\text{H}^+]/[\text{B}] \quad (4)$$

Usually HEPES is used in large enough quantities as a buffer[64,65] for biological systems to maintain constant pH during experiments. In our case, however, the largest amounts of HEPES are comparable to the maximum initial substrate concentration. This means that in the beginning of reaction or for small substrate concentrations, i.e., close to the logic-**0**, HEPES will operate as a buffer and keep the pH of the solution at a constant, initially titrated level. When more acid is produced in the enzymatic reaction (1), the buffering capacity of HEPES will eventually be overwhelmed, and the pH will rapidly decrease and stay constant at the level determined by $K_A$ and the final amount of the produced butyric acid. As a result, an inflection region appears on the curve of pH vs. the initial substrate concentration which thus becomes sigmoidal. Note that the presence of a buffer significantly modifies the system's response near logic-**0** of the input, whereas the response curve near logic-**1** is typically sufficiently "flat" anyway, as described earlier.[50,55]

However, additional challenges arise in this approach. Ideally, we would like to have such an amount of HEPES that the "flat" low-noise regions extent far around both logic points.[52,55] However, if there is too little HEPES, then it will be quickly overwhelmed even by small amounts of butyric acid so that the small-slope region at logic-**0** will be very narrow. On the other hand, when the amount of HEPES is large, it will act as a buffer throughout the entire range of input concentrations so that the pH at logic-**1** will be close to that at logic-**0**, and it may become difficult to distinguish between the outputs at those two logic points. Furthermore, too much HEPES can increase the slope at logic-**1**, because the reaction will be slowed down and away from saturation. Thus, the problem of establishing the optimal concentration of the buffer for adequate suppression of the analog noise over the broad range of the logic input concentrations around each logic point requires numerical optimization based on the initial data fits. This is illustrated in Section 5.

Figure 2 exemplifies the filtering effect by showing sets of data taken for increasing amounts of HEPES. While the onset of the sigmoidal behavior is clearly seen, the plots also illustrate that the data are noisy, as typical for such enzymatic systems. Figure 3 shows the pH dependence on the reaction time $t$ and initial substrate concentration, for 100 mM of HEPES. For later times ($t > 20$ min) and large enough ethyl butyrate concentrations (over approximately 20 mM), the pH decrease slows down, resulting in a flat region, which illustrates the saturation mentioned earlier, typical of enzymatic processes. Similar trends in the pH dependence were observed also for other studied concentrations of HEPES. We note that the experimental data here are also noisy, which underscores the importance of parameter selection for the filtering effect in such a way that good separation is maintained between the "logic" reference values.

## 4.     Kinetic Equations and Noise Analysis

In order to analyze our data, exemplified in Figures 2 and 3, we use a kinetic rate-equation description of the processes involved. We already noted the complexity of the process steps involved. For a realistic description, we therefore have to limit the number of fitted rate constants utilized. This has been a standard practice in such applications because we seek a semi-quantative overall description of the response surface of the process, with the detailed behavior relevant for signal handling only near the logic-point values. Thus, we use a simplified kinetic description of reaction (1) assuming no reversibility in the formation of the intermediate complex. Indeed, the available kinetic data[60] for this type of reaction suggest that, in our regime of the parameters, for most of our pH range (except perhaps in the far saturation regime), for the reactant concentrations used, the rate of the reverse reaction will be negligibly small. Furthermore, we use a description whereby the rate equations for hydrolysis process (1), are written in terms of the total (active and acidified) enzyme and complex concentrations, whereas the reduction in the process rates due to the acidification equilibria is lumped in the effective rate parameters $R'(t)$ and $r'(t)$, which replace $R$ and $r$ in (1): We have

$$\frac{d[S]}{dt} = -R'(t)[S]\{E_0 - [C]\}$$
$$\frac{d[C]}{dt} = R'(t)[S]\{E_0 - [C]\} - r'(t)[C] \quad (5)$$
$$\frac{d[P]}{dt} = r'(t)[C]$$

where $E_0$ is the total initial amount of the enzyme, and, as just mentioned, we incorporate the acidification eqilibria shown in (3) in the effective rate parameters by defining

$$R'(t) = \frac{K_E}{[H^+](t) + K_E} R$$
$$r'(t) = \frac{K_C}{[H^+](t) + K_C} r \quad (6)$$

The concentration $[H^+](t)$ in (6), which yields the pH value, is determined from the charge balance equation. Since the enzyme and complex concentrations are very small compared to other chemicals, we can neglect their contributions. The resulting equation is

$$[H^+](t) = [H^+]_0 + \frac{K_A}{K_A + [H^+](t)}[P](t) + \frac{K_{HEPES}}{K_{HEPES} + [H^+](t)}[HEPES] - [B^-]_0 \quad (7)$$

where $[H^+]_0 = 10^{-7}$ M is the concentration corresponding to the initial titration of the system to $pH_0 = 7$; $[B^-]_0$ denotes the initial concentration of the HEPES anions after its instantaneous dissociation equilibration (at $pH_0 = 7$), whereas the total amount of HEPES introduced has been, as a concentration, denoted by [HEPES] already in the captions Figures 1–3. As a result of a numerical solution of the various coupled equations introduced, we can obtain the dependence of the pH($t$) on the reaction time.

As alluded to earlier, there are several sources of noise[11,52,55] in biochemical information processing at the level of network elements, as well as at the level of the network as a whole. The buildup of noise must be avoided to allow scalability and fault tolerance. The former, single-element (gate) noise results, for instance, from the inaccuracy of the logic function itself, as seen in Figures 2 and 3, and reflects fluctuations in the chemical concentrations and other experimental conditions and parameters. Our aim is to design network elements that will minimize such analog noise *amplification* as the signal is processed. Other types of noise, specifically, digital noise, are handled at the network-design level, not addressed here. Filtering is the primary tool for avoiding amplification of analog noise. This property of a filter can be quantified as follows.

We define dimensionless logic-range input ($x$) and output ($y$) variables, encountered earlier (see the inset in Figure 1):

$$x = \frac{[S](t=0)}{[S]_{max}(t=0)} \quad (8)$$

$$y = \frac{pH_0 - pH(t)}{pH_0 - pH_{[s]_{max}}(t)} = F(x) \quad (9)$$

where $[S]_{max}(t=0)$ is the maximum initial concentration (here 100 mM) of the input substrate in our experiments, selected as logic-**1**; $pH_0$ was defined earlier (=7); and $pH_{[s]_{max}}(t)$ is the output value at the reference time $t$ for the logic-**1** input, i.e., for $[S]_{max}(t=0)$, which thus defines the logic-**1** of the output. Thus, the gate-response function $y = F(x)$, for our trivial, "identity" gate connects the logic points at $(x,y) = (0,0)$ and $(1,1)$.; see Figure 1 inset.

The response function (9) can be calculated from the presented solution once four adjustable parameters, $R, r, K_E, K_C$ are first determined from a least-squares fit of the experimentally measured pH data, as reported in the next section. Other quantities and parameters are known or were taken from the literature; the latter were the tabulated values for $pK_A = 4.83$ and $pK_{HEPES} = 7.55$, which were validated by performing control experiments in which butyric acid and HEPES were mixed directly.

With the response function $F(x)$ calculated, we perform numerical analysis[48,50] of our logic filter to gauge its noise amplification/suppression properties in the vicinity of the two logic points, **0** and **1**. We define the noise amplification factor as the ratio of the maximum of the two output noise distribution spreads, $\sigma_i^{out}$ (computed for each logic point, $i = 0, 1$) and the input noise spread, $\max(\sigma_i^{out}/\sigma^{in})$. Here the spread of the output distribution is defined as the root-mean-square width, for instance,

$$\sigma_i^{out} = [\langle y^2 \rangle_i - \langle y \rangle_i^2]^{1/2} \quad (10)$$

with the averages $\langle \cdots \rangle$ at each logic point are computed with respect to the input noise distribution function which is assumed to be Gaussian with the same variance, $\sigma^{in}$, at both $i = 0, 1$. (Actually, half-Gaussian for a logic point with only positive signal values physically possible, such as our logic-**0** input, [S] = 0.) This ratio allows us to determine how large are the deviations in the output signal as compared to the assumed spread in the input signal. For a good filter, $\max(\sigma_i^{out}/\sigma^{in})$ should be less than 1 (means, noise suppression) for typical input signal spread values in biochemical systems, which have $\sigma^{in}$ at least a couple of percent on the scale of the logic-interval range of 1.

5.  **Results**

Experimentally measured dependence of the pH on the reaction time $t$ and initial ethyl butyrate (substrate) concentration is illustrated in Figures 2 and 3. For a specific parameter extraction we used the full time-dependent data sets taken at [HEPES] = 100 mM, which were of the best detail and quality, yielding the estimates $R = 60.0 \pm 0.2$ (mM·s)$^{-1}$, $r = 26.9 \pm 0.7$ s$^{-1}$, $K_E = 0.00680 \pm 0.00003$ mM, and $K_C = 0.020 \pm 0.006$ mM. The quality of the least-squares fits was quite good, and these parameter values were consistent with other data sets taken. Note that $K_C \gg K_E$, similar to the results found earlier with methyl-*n*-butyrate as the substrate for esterase.[58,60]

The onset of the sigmoidal profile in the response curves can be seen in Figure 2, and is also clearly present in the data shown in Figure 3. Specifically, Figure 2 shows how pH of the system depends on the substrate concentration at a fixed reaction time, 120 min, for increasing amounts of HEPES: 0, 50, 100 mM. As expected, with larger concentrations of HEPES the response curves around logic-**0** point become more "flat" whereas the slopes at logic-**1** slightly increase. This is also seen in the inset in Figure 1, where the same curves are shown in terms of the logic-range variables, $y$ vs. $x$. A tradeoff involved in getting the filtering effect is the trend, clearly visible in Figure 2, of decreasing the difference between the pH values at the two logic points, $\Delta \text{pH}_{01} = \text{pH}_0 - \text{pH}_1$. This occurs because the pH-drop signal buildup is delayed by the filtering effect. Without HEPES we have $\Delta \text{pH}_{01} \sim 4.1$, but this difference drops to $\sim 2.7$ for [HEPES] = 100 mM. This property may make it more difficult to distinguish between the outputs (pH values) at logic-**0** and **1** in real-life applications, due to another source of noise, obvious in Figure 2 and 3: the intrinsic noise in the logic-element functioning itself (whereas filtering is aimed at reducing noise *amplification* from input to output).

The numerically evaluated dependence of the maximum (over the two logic-point values) noise amplification factor, $\max(\sigma_i^{out}/\sigma^{in})$, vs. reaction time $t$ and HEPES concentration is shown in Figure 4, assuming a rather large input noise spread, $\sigma^{in} = 0.3$ (30%). The dots correspond to the experimental conditions at which data shown in Figure 2 were obtained. With increasing time and HEPES concentration, $\max(\sigma_i^{out}/\sigma^{in})$ decreases until it reaches its minimum, at $t \sim 70 - 160$ min, [HEPES] $\sim 250$ mM (this region is marked in the figure). Note that in this region we would have $\max(\sigma_i^{out}/\sigma^{in}) \sim 0.7 < 1$, which would yield actual noise suppression. However, this value is still not vanishingly small because at $\sigma^{in} = 0.3$ there is a noticeable contribution to the output signal from the parts of the response curve with large slope (close to the inflection). For smaller input-noise spread (than the assumed 30%) the filtering effect will be somewhat better, but the overall results and trends are qualitatively similar.

The region of the optimal noise suppression marked in Figure 4, is interesting for potential applications because in addition to a small noise amplification factor (~ 0.7), the pH variation between the two logic points is not much reduced: $\Delta pH_{01}$ is ~ 2 or larger; see Figure 5. We point out that, in Figure 4 there is another range of the reaction times and HEPES concentrations for which $\max(\sigma_i^{out}/\sigma^{in}) < 1$. It is located at $t \sim 50$ min and [HEPES] > 300 mM. However, as can be seen from Figure 5, in this region $\Delta pH_{01}$ is somewhat smaller than ~ 1, which makes this range of parameters less favorable for filter operation.

Near the logic-**0** point, when pH deviates only a little from its initial titration value, $pH_0$, we can ignore the effect of the variation of pH on the enzymatic kinetics, represented by (6) in our model. Then the enzymatic kinetics yielding its output product $P(t)$, and the filtering effect of the buffer, the latter represented in the degree to which $P(t) > 0$ can decrease the pH, decouple. One can thus show that the presence of the buffer decreases the absolute value of the (negative) slope of pH as a function of the input, [S], at the logic-**0** (i.e., at [S] = 0) as follows:

$$\frac{\left(\frac{d\text{pH}}{d[S]}\Big|_0\right)_{[\text{HEPES}]>0}}{\left(\frac{d\text{pH}}{d[S]}\Big|_0\right)_{[\text{HEPES}]=0}} \approx \frac{1}{1 + \frac{\frac{K_{\text{HEPES}}}{[H^+]_0}}{\left(\frac{K_{\text{HEPES}}}{[H^+]_0} + 1\right)^2} \cdot \frac{[\text{HEPES}]}{[H^+]_0}} \quad (11)$$

From this expression we see not only that increasing buffer concentration leads to a smaller slope and consequently, lesser noise amplification, as expected, but also that both the initial pH and the buffer dissociation constant affect filter performance. For different initial pH values, a buffer is preferable such that (the negative decimal logarithm of) its dissociation constant is very approximately equal to $\text{pH}_0$. This maximizes the coefficient of $[\text{HEPES}]/[H^+]_0$ in the denominator, yielding the optimal analog noise handling at logic-**0**. Note that this condition is relatively well met by our system: $pK_{\text{HEPES}} = 7.55 \sim \text{pH}_0 = 7$.

## 6. Conclusion

In this work we experimentally demonstrated a biochemical logic filter with artificially induced sigmoid pH-drop response. We used esterase-catalyzed hydrolysis of ethyl butyrate as the substrate. By performing this reaction with the HEPES buffer supplied in measured amounts, we were able to suppress the change in the pH at small initial concentrations of the substrate. As a result, the response curve of the filter (dependence of the pH on substrate concentration) changes to the sigmoid one for which analog noise can in principle be suppressed around both logic points provided that we have the correct amount of HEPES in our system and properly select other system parameters as discussed in Section 4.

We performed kinetic modeling of the system and determined the optimal required amount of HEPES and reaction times at which maximum noise suppression occurs. For this, we first numerically fitted the experimental data to the solution of the system of kinetic and charge balance equations, thus fixing the unknown reaction parameters. With these quantities known, we were able to study general noise properties of this system as functions of the HEPES concentration and reaction time. We found that the optimal amount of HEPES at which maximum suppression of noise would occur is $\sim 250$ mM, with reaction times $t \sim 70 - 160$ min. While our reaction times were in this range, we only went up to 100 mM of HEPES. This already yielded a significant improvement (means, reduction) in the analog noise amplification factor, as indicated by the values reported in Figure 4. We also found that in general, optimum performance of the filter is possible only provided one works with systems such that the initial pH is close to (the negative decimal logarithm of) the dissociation constant of the selected weak-acid buffer. The method as described, is easy to realize in practice and shows promise for use in information processing networks, because no cross-reactions are introduced: the buffer acid is typically not a part of the (bio)chemical processes in the system.

**Acknowledgements**

The authors thank Dr. J. Halámek for helpful scientific input and collaboration. The work in Spain (M.P.) was funded by Ramón y Cajal program, MICINN. We acknowledge research funding by the US-NSF, under grant CCF-1015983.

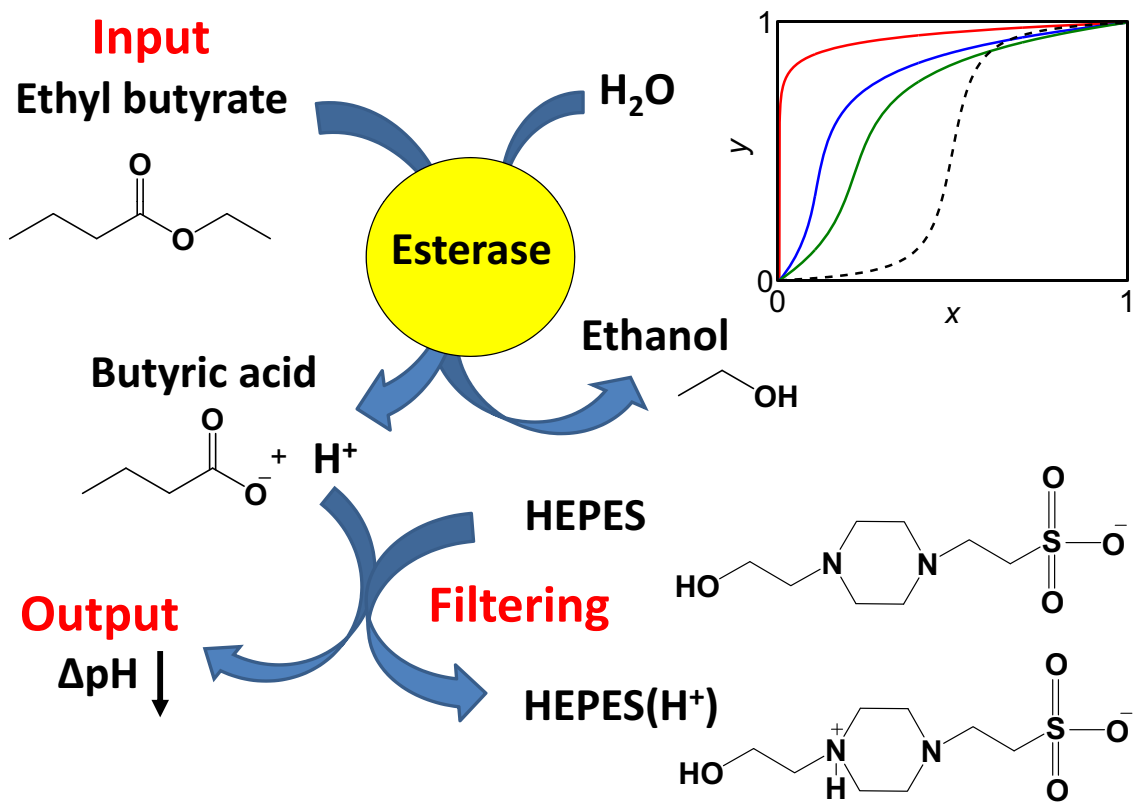

**Figure 1.** Schematic presentation of the buffering-based pH-signal "logic filter." The reaction biocatalyzed by an enzyme, here esterase, results in the hydrolysis of ethyl butyrate (the logic *Input*) to yield butyric acid which releases $H^+$ ions upon dissociation. A measured quantity of a buffer, here HEPES, if introduced, consumes most of the biocatalytically produced $H^+$ ions when the input is applied at a low concentration. The pH change (the logic *Output*, measured by the pH drop, as indicated by an arrow) sets in when the biocatalytically produced $H^+$ ions overwhelm the buffer. The biocatalytic process and buffering combined, yield a sigmoidal dependence of the pH change as a function of the input concentration. The inset illustrates the onset of the sigmoidal response in our experimental system. The solid curves show the output, *y*, vs. the input, *x*, properly redefined/rescaled to vary in the "binary-logic ranges" from 0 to 1, as explained in the text. Experimental data were fitted by using rate equations appropriate for the processes involved, and the results are shown, here for the reaction time 120 min, for increasing buffer (HEPES) concentrations. The top (red) curve corresponds to [HEPES] = 0; middle (blue): [HEPES] = 50 mM, bottom (green): [HEPES] = 100 mM. The dashed black curve does not correspond to experimental data but rather illustrates a desirable, "ideal" filter response with small slopes at both binary logic points **0** and **1**, and with a steep, symmetrically positioned inflection region in the middle.

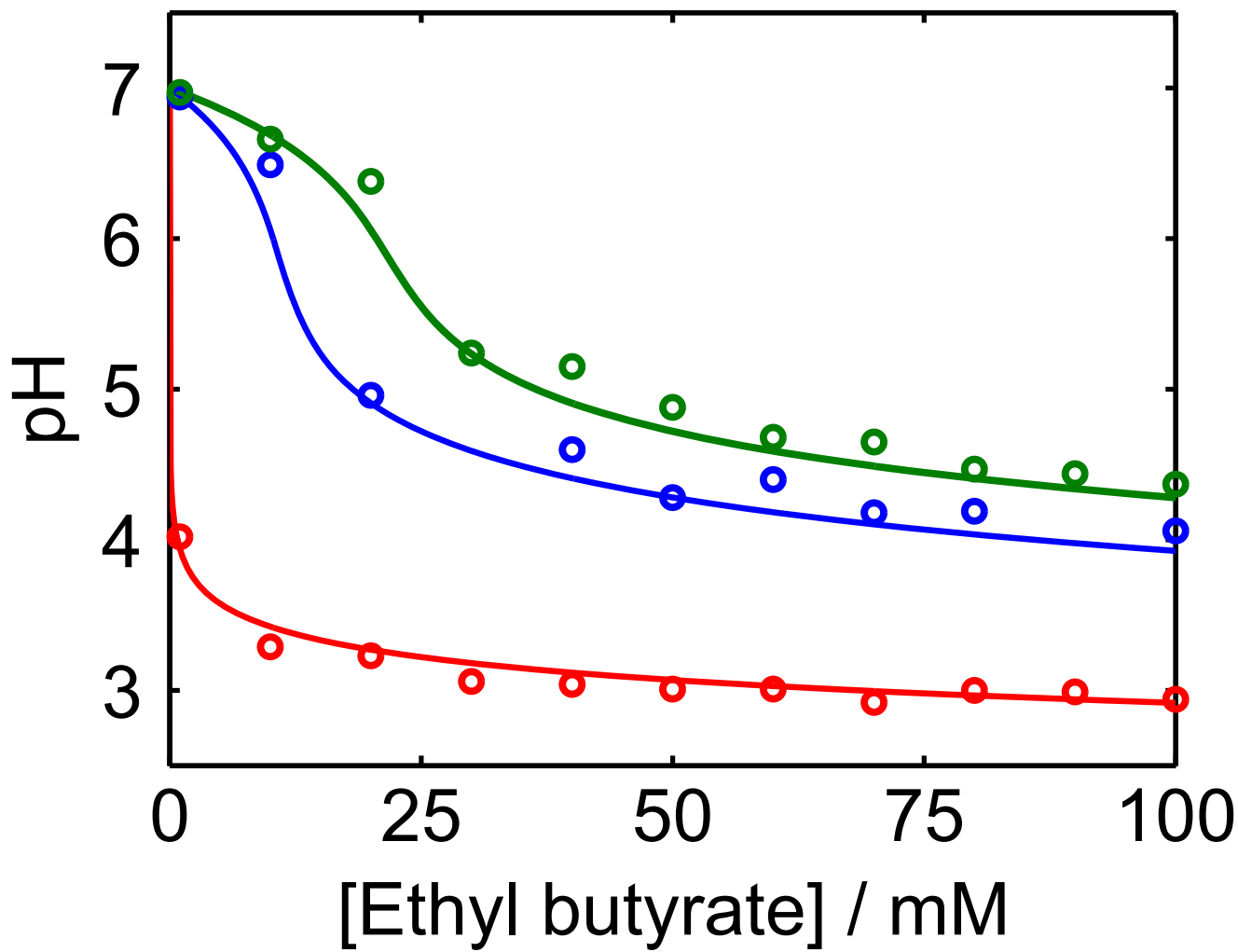

**Figure 2**. Measured pH values at the reaction time $t = 120$ min, shown vs. the initial substrate concentration, for different amounts of HEPES. Red (bottom) symbols/curve correspond to [HEPES] = 0, blue (middle): [HEPES] = 50 mM, green (top): [HEPES] = 100 mM. The circular symbols are the actual pH values, whereas the solid curves are the theoretical model fits. (These curves were shown in the inset in Figure 1, rescaled in terms of the logic-range variables).

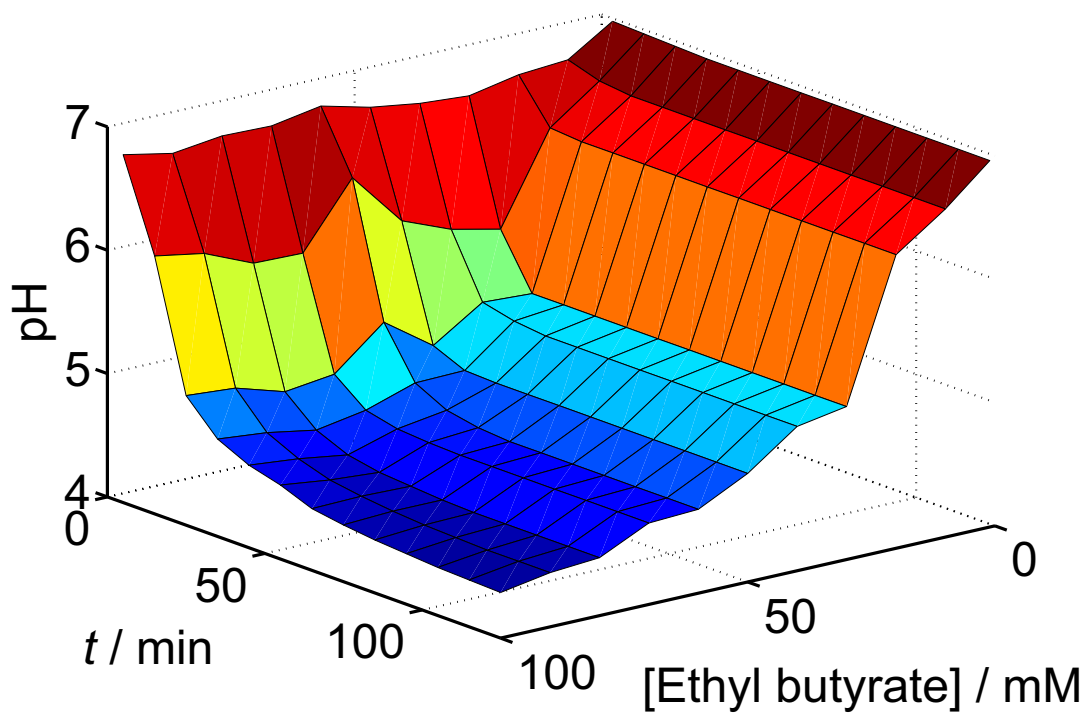

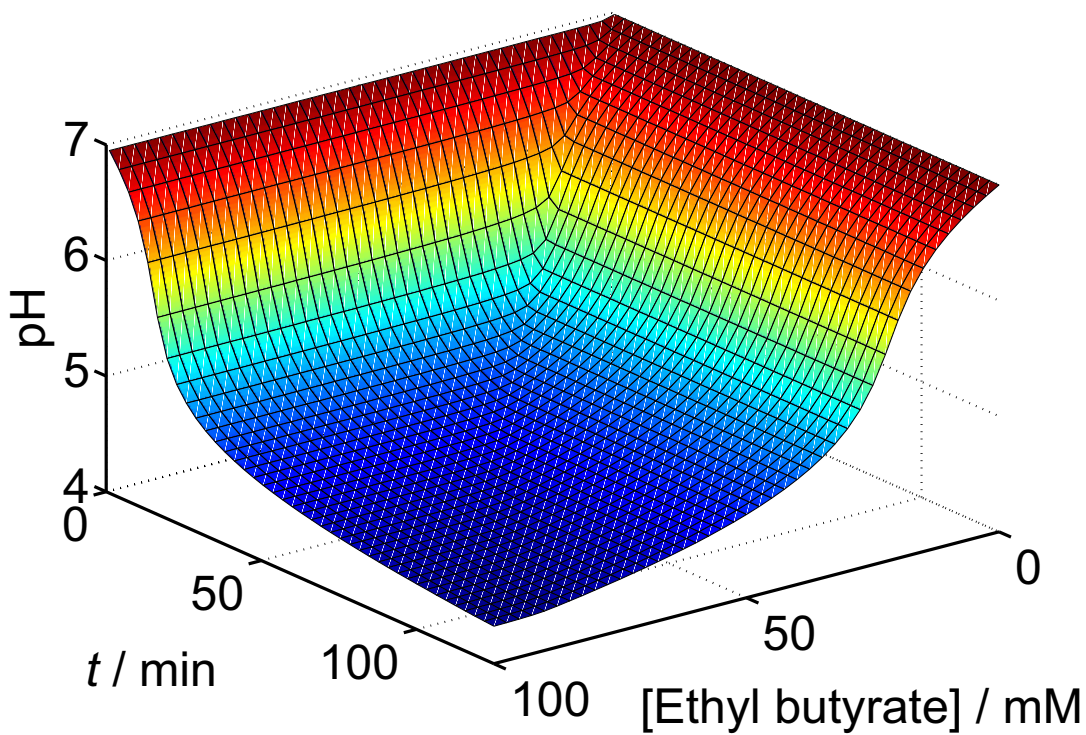

**Figure 3.** Top: Experimental dependence of pH on the initial substrate concentration (ethyl butyrate) and reaction time, for [HEPES] = 100 mM. Bottom: Numerically computed dependence for this system, based on the kinetic model.

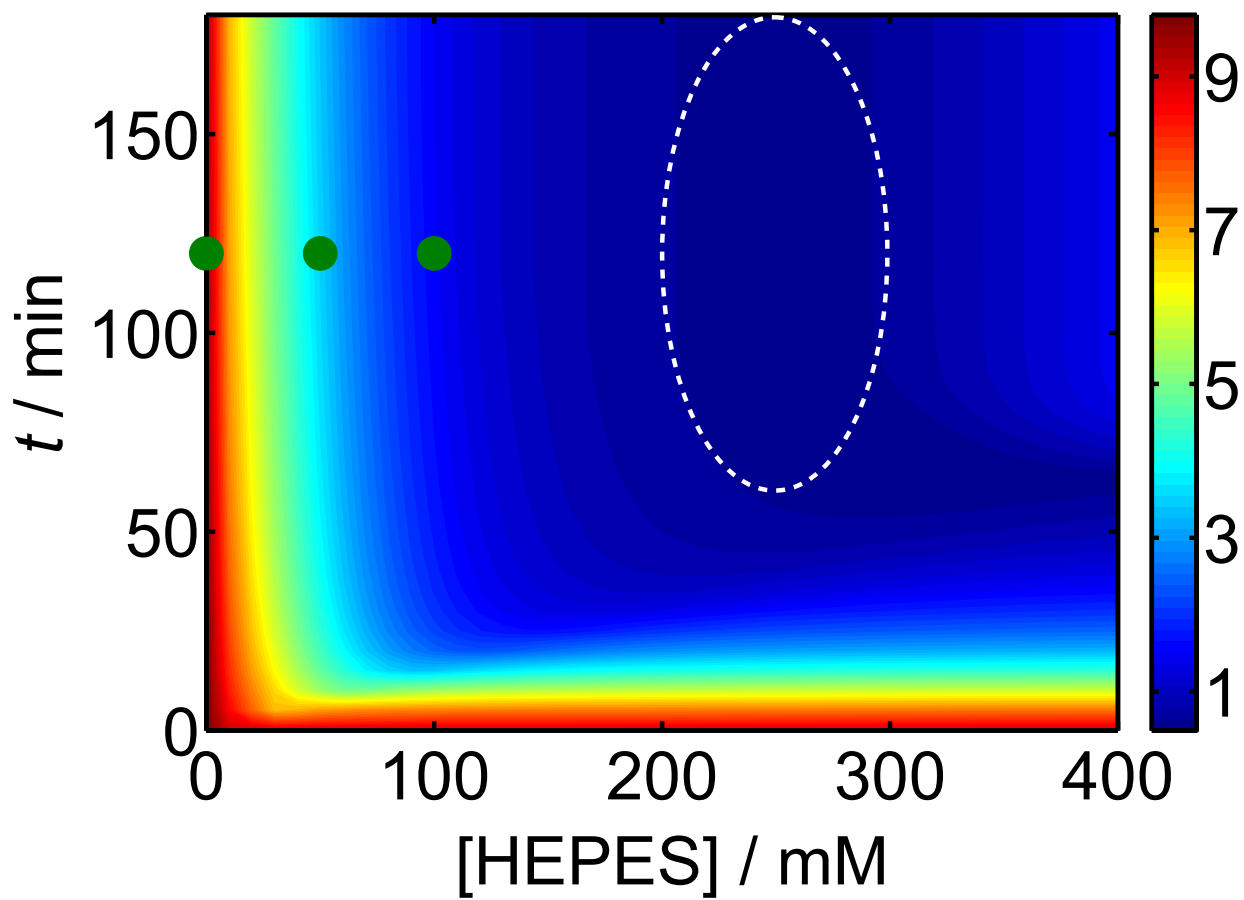

**Figure 4**. Color-coded contour plots of the noise amplification factor $\max(\sigma_i^{out}/\sigma^{in})$ as a function of the concentration [HEPES] and reaction time $t$. The dots mark the conditions corresponding to the curves in Figure 2, with the corresponding noise amplification factor values 9.45, 3.52, 1.56, for [HEPES] = 0, 50, 100 mM, respectively. The broken-line ellipse encircles the optimal-parameter region of filter operation for which the strongest suppression of analog noise would be possible.

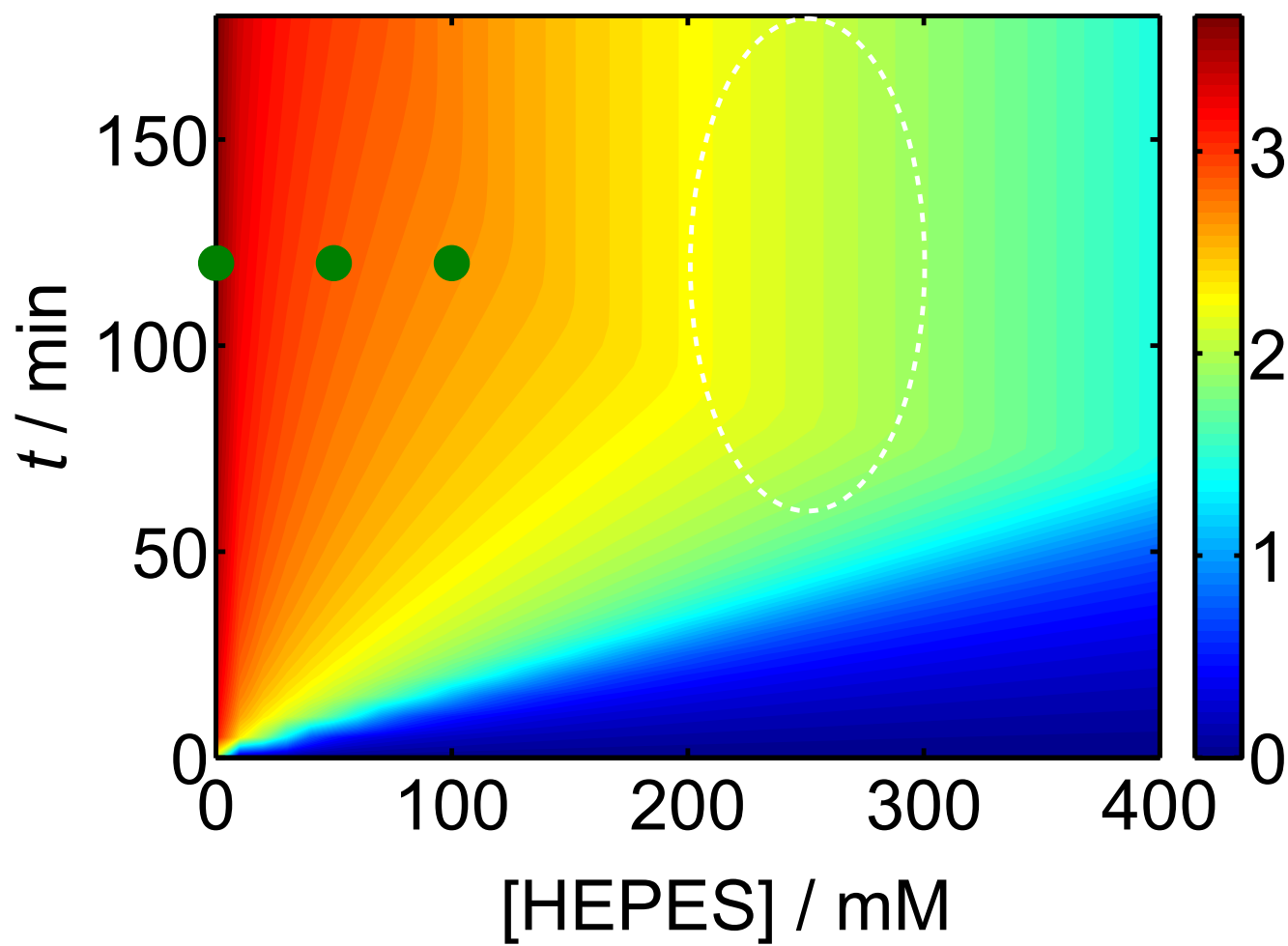

**Figure 5.** Color-coded contour plots of $\Delta pH_{01}$. All axes, notation, and markings are the same as in Figure 4.

**Graphical Abstract (Table of Contents Image)**

Biochemical logic filter with buffer-induced sigmoid pH-drop response of an enzyme-catalyzed reaction is experimentally realized and optimized by kinetic modeling.

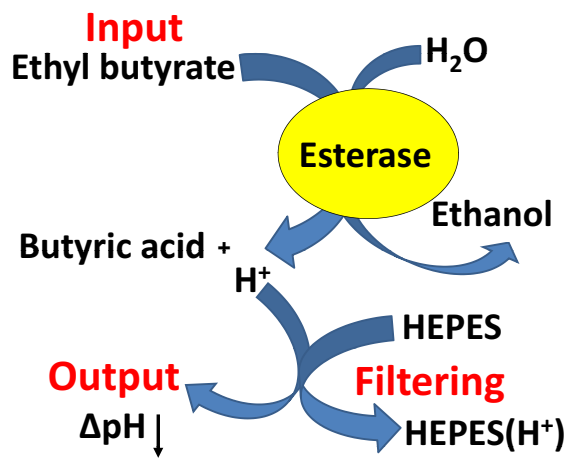
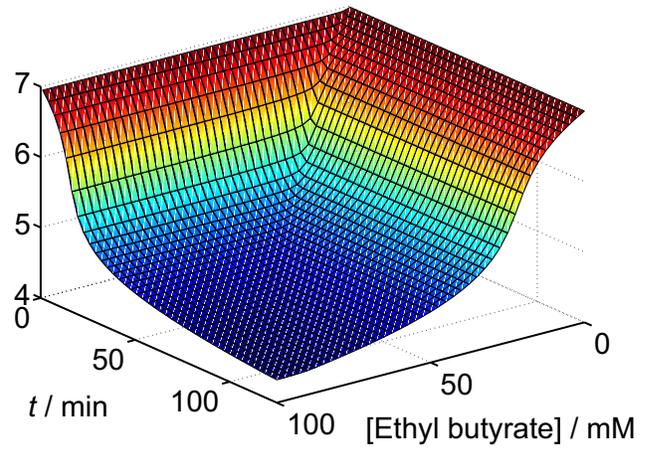